\documentclass[aps,prc,reprint,superscriptaddress,showpacs,amsmath,amssymb,letter]{revtex4-1}

\usepackage{graphicx}
\usepackage{bm}
\usepackage{color}

\begin{document}

\title{Neutron degeneracy and plasma physics effects on radiative
neutron captures in neutron star crust}

\author{P.~S.~Shternin}\email{pshternin@gmail.com}
\affiliation{Ioffe Physical-Technical Institute, Politekhnicheskaya
26, 194021 St.-Petersburg, Russia}
\affiliation{St.-Petersburg State Polytechnical University,
Politekhnicheskaya 29, St.-Petersburg 195251, Russia}

\author{M.~Beard}
\author{M.~Wiescher}
\affiliation{Department of Physics and The Joint Institute for
Nuclear Astrophysics, University of Notre Dame, Notre Dame,
Indiana 46556, USA}

\author{D.~G.~Yakovlev}
\affiliation{Ioffe Physical-Technical Institute,
Politekhnicheskaya 26, St.-Petersburg 194021, Russia}
\affiliation{St.-Petersburg State Polytechnical University,
Politekhnicheskaya 29, St.-Petersburg 195251, Russia}
\affiliation{Department of Physics \& The Joint Institute for Nuclear
Astrophysics, University of Notre Dame, Notre Dame, Indiana 46556,
USA}

\date{\today}

\begin{abstract}
We consider the astrophysical reaction rates for radiative neutron
capture reactions ($n,\gamma$) in the crust of a neutron star. The
presence of degenerate neutrons at high densities (mainly in the
inner crust) can drastically affect the reaction rates. Standard
rates assuming a Maxwell-Boltzmann distribution for neutrons can
underestimate the rates by several orders of magnitude. We derive
simple analytical expressions for reaction rates at a variety of
conditions with account for neutron degeneracy. We also discuss
the plasma effects on the outgoing radiative transition channel in
neutron radiative capture reactions and show that these effects
can also increase the reaction rates by a few orders of magnitude.
In addition, using detailed balance, we analyze the effects of
neutron degeneracy and plasma physics on reverse ($\gamma,n$)
photodisintegration. We discuss the dependence of the reaction
rates on temperature and neutron chemical potential and outline
the efficiency of these reactions in the neutron star crust.
\end{abstract}


\maketitle

\section{Introduction}
\label{S:intro}

Nuclear reactions in the atmosphere and the crust of accreting
neutron stars affect important observational manifestations such
as X-ray bursts and superbursts (e.g., Refs.\
\cite{sb06,schatz03,cummingetal05,gu07}) as well as deep crustal
heating of neutron stars in X-ray transients (e.g., Refs.\
\cite{hz90a,hz03,bbr98,gu07}). In the vicinity of the neutron drip
density ($\rho\sim 4\times 10^{11}$~g~cm$^{-3}$ for the
cold-catalyzed crust and $\rho\sim 6\times 10^{11}$~g~cm$^{-3}$
for the accreted crust \cite{hz90a}) and beyond in the inner crust
the dense matter contains an increasing amount of free degenerate
neutrons (see, e.g., Ref.\ \cite{hpyBOOK}). Neutron capture and
reverse reactions are important components of nuclear burning
under these conditions \cite{gkm08}. Standard thermonuclear
neutron capture rates, which are used in reaction network
simulations of nucleosynthesis in stars or supernova explosions,
are obtained (e.g., Ref.\ \cite{rt00}), assuming the classical
Maxwell-Boltzmann distribution of neutrons.  However, free
neutrons in the neutron star crust can be degenerate, in
particular when the density exceeds the neutron drip
point~\cite{hpyBOOK}. For instance, ground-state (cold-catalyzed)
matter at $\rho=6.2 \times 10^{12}$~g~cm$^{-3}$ has a neutron
Fermi energy of $\approx$ 2.6~MeV \cite{nv73}. Consequently
neutron degeneracy needs to be taken into account for neutron
capture rates under such conditions.

In addition, the dense stellar plasma of the neutron star crust
strongly affects emission, absorption, and propagation of photons
\cite{sy09} and therefore modifies radiative capture and
photodisintegration reactions, like ($n,\gamma$) and ($\gamma,n$).
Because of the high density, the electron plasma frequency
$\omega_\mathrm{p}$ can be of the order of or higher than
characteristic frequencies of radiative transitions in nuclei.
Under these conditions, well-defined elementary electromagnetic
excitations (photons or plasmons) become either suppressed or
forbidden (e.g., Ref.~\cite{abr84eng}) although radiative
transitions are not suppressed  because they can be realized by
emission (or absorption) of excess energy to (from) the plasma as
a collective system \cite{sy09}. These plasma physics effects can
be important since they may enhance the radiative transition
strength.

In  Sec.\ \ref{S:deg_n} we discuss the effects of neutron
degeneracy on ($n,\gamma$) radiative neutron capture reactions in
dense matter. In Sec.\ \ref{S:plasma} we analyze plasma effects on
the outgoing radiative transition channel of ($n,\gamma$)
reactions. In Sec.\ \ref{S:reverse} we consider the same neutron
degeneracy and plasma physics effects on inverse ($\gamma,n$)
photodisintegration reactions. We discuss our results in Sec.\
\ref{S:discuss} and summarize them in Sec.\ \ref{S:concl}. For
brevity, we use the units in which the Boltzmann constant $k_B=1$.

\section{Reaction rates for degenerate neutrons}
\label{S:deg_n}

We start with the outline of the ($n,\gamma$) radiative capture
rates in stellar environments (e.g., Ref.\ \cite{rt00}). Let the
cross section $\sigma_{ab}(E)$ refer to the reaction $X^{(a)}+n\to
Y^{(b)}+\gamma$, where $a$ and $b$ label different energy levels
of a target nucleus $X$ and a resultant nucleus $Y$, respectively,
and $E$ is the center-of-mass energy of the reactants. In stellar
matter at local thermodynamic equilibrium, the total cross section
$\sigma^*(E)$ of the reaction $X+n\to Y+\gamma$ includes neutron
capture on the ground state and all thermally populated states,
\begin{equation}\label{eq:sigma_astro}
  \sigma^*(E)=\frac{\sum_a g_a \exp\left(-E^{(a)}_{\mathit{X}}/T\right)
  \sum_b \sigma_{ab}(E)}{\sum_a g_a
  \exp\left(-E^{(a)}_{\mathit{X}}/T\right)},
\end{equation}
where $E^{(a)}_\mathit{X}$ is the energy of  level $a$ and $g_a$
is its statistical weight. The summation in the denominator
normalizes the distribution of target nuclei over the energy
levels $a$ (gives the internal partition function of the target
nucleus). Asterisk * means that thermally excited nuclear levels
in stellar matter are included.

The astrophysical reaction rate contains the average $\langle
\sigma^* v\rangle$ of the total cross section with the energy
distribution $f(E)$ of the interacting particles. For
nonrelativistic reactants (considered in this paper) the collision
energy is $E=\mu v^2/2$, where $\mu$ is the reduced mass (very
close to the neutron mass $m_n$), and $v$ is the relative velocity
of a neutron with respect to nucleus at large separations. Then
\begin{equation}
  \label{eq:rate}
  \langle \sigma^* v\rangle = \sqrt{\frac{2}{\mu}} \frac{1}{\cal N}
  \int_0^\infty E\sigma^*(E) f(E)\, \text{d} E,
\end{equation}
where ${\cal N}$ is the normalization factor
\begin{equation}
  {\cal N}= \int_0^\infty \sqrt{E}\, f(E)\, \text{d}E.
\label{eq:N}
\end{equation}
We call $\langle \sigma^* v\rangle$ [cm$^3$~s$^{-1}$] the reaction
rate coefficient. The rate itself (for instance, per unit volume,
cm$^{-3}$~s$^{-1}$) is $n_X n_n \langle \sigma^* v\rangle$, where
$n_n$ and $n_X$ are number densities of neutrons and reacting
nuclei, respectively.

Astrophysical reaction rates at typical stellar temperatures are
based on a Maxwell-Boltzmann distribution of the particles,
$f_{\text{MB}}(E)=\exp(-E/T)$. At high densities in the neutron
star crust, neutrons can become degenerate, which modifies the
reaction rate. At these conditions the nuclei are not freely
moving particles but are confined in a strongly coupled Coulomb
liquid or a Coulomb crystal (e.g., Ref.\ \cite{hpyBOOK}). Because
the neutrons are much lighter than the nuclei, the energy
distribution function $f(E)$ in Eq.~(\ref{eq:rate}) can be
approximated by a Fermi-Dirac distribution
$f_{\text{FD}}(E)=\left[1+\exp((E-\mu_n)/T)\right]^{-1}$, where
$\mu_n$ is the neutron chemical potential. In this approximation
we neglect recoil effects and nucleus motion. In the inner crust
of the neutron star the nuclei can be bulky and occupy a
non-negligible fraction of volume \cite{st83}. Here we employ the
model of a free neutron gas with local number density $n_n$ which
occupies the space between the nuclei. Though this model is rather
accurate near the neutron drip point, it becomes less accurate at
higher densities where free neutrons constitute a strongly
interacting Fermi liquid \cite{nv73}.

Let the average $\langle\sigma^* v\rangle_{\text{MB}}$ be obtained
with the Maxwell-Boltzmann distribution and $\langle \sigma^*
v\rangle_{\text{FD}}$ be calculated with the Fermi-Dirac
distribution. Many calculated reaction rate coefficients
$\langle\sigma^* v\rangle_{\text{MB}}$ for neutron capture
reactions are available in the literature (e.g., Ref.\
\cite{Cyburtetal10} and references therein). In a neutron star
crust $\langle \sigma^* v\rangle_{\text{FD}}$ depends on $T$ and
$\mu_n$ (or, equivalently, on $T$ and $n_n$). For practical
applications we introduce the ratio
\begin{equation}\label{eq:R}
  R_n\equiv\frac{\langle\sigma^*
  v\rangle_{\text{FD}}}{\langle\sigma^*
  v\rangle_{\text{MB}}}.
\end{equation}
These ratios are easier to calculate and approximate than $\langle
\sigma^* v\rangle_{\text{FD}}$; the derivation of these ratios ise
the main subject of the present paper.

Generally, accurate calculations of $\langle \sigma^*
v\rangle_{\text{FD}}$ and $R_n$ require the cross sections
$\sigma_{ab}(E)$ obtained from experiment or nuclear reaction
codes. Detailed calculations of $\langle
\sigma^*v\rangle_{\text{FD}}$ would be a valuable project for the
future.  Here we restrict ourselves to a simplified approach. It
allows us to demonstrate the importance of the effects of neutron
degeneracy and is sufficiently accurate for a wide range of
temperatures and densities. First, we neglect the contribution of
thermally excited states (setting thus $\sigma^*(E)=\sigma(E)$).
This is a valid approach if the energy of the first excited level
of the target nucleus is higher than the temperature in the
neutron star crust ($T \lesssim 2 \times 10^9$ K$\approx 0.2$
MeV). For threshold (endothermic) reactions, it should also be
higher than the reaction threshold $E_0$. Second, we note that the
reaction rates at low temperatures correspond to cross sections
$\sigma(E)$, which are characterized by typical power-law behavior
that we therefore adopt in our analysis:
\begin{equation}
\label{eq:sigma_pl}
  \sigma(E) = \sigma_a \,(E-E_0)^\nu \quad {\rm at}~~E_0\leq E \lesssim E_\mathrm{max}.
\end{equation}
Here, $\nu$ is a power-law index, $\sigma_a$ is a normalization
constant, $E_0$ is a reaction threshold (with $E_0=0$ for
exothermic reactions), and $E_\mathrm{max}$ is the maximum energy
up to which the approximation (\ref{eq:sigma_pl}) holds. We treat
$E_0$, $\sigma_a$, $\nu$, and $E_\mathrm{max}$ as input
parameters. For a given reaction, they can be adopted from a
nuclear database or calculated using a nuclear reaction code. In
our approximation, the factor $R_n$ depends on $T$, $\mu_n$,
$\nu$, and $E_0$; the parameter $\sigma_a$ cancels out in the
ratio (\ref{eq:R}); $E_\mathrm{max}$ is required to check the
validity of the calculated $R_n$ for given conditions.

The reaction rates are strongly affected by the neutron energy
distribution. Typical energies of nondegenerate neutrons are $E
\lesssim T$. In a strongly degenerate gas ($\mu_n \gg T$) the
majority of neutrons belong to the Fermi sea and have much higher
energies $T\ll E \lesssim \mu_n$. In this case, there is also a
smaller (but non-negligible) amount of neutrons, with energies
above $\mu_n$: $\mu_n\lesssim E\lesssim \mu_n+T$. Their
distribution $f_\mathrm{FD}(E) \approx
\exp((\mu_n-E)/T)=\exp(\mu_n/T)\,f_\mathrm{MB}(E)$ is close to
Maxwellian and represents the Maxwellian tail of the Fermi-Dirac
distribution.  In the following we demonstrate that these
different energy ranges of $f_\mathrm{FD}(E)$ correspond to
different neutron capture regimes.

For all nuclear reactions shown in the paper we use the cross
sections obtained with the statistical model Hauser-Feshbach (HF)
code TALYS-1.2 \cite{Talys10}. Statistical model theory
\cite{hf52} uses the concept of averaged transmission coefficients
to describe the formation and subsequent  decay of a compound
nucleus formed after a projectile impinges on a target nucleus. In
this scenario the reaction sequence for neutron capture becomes
$X^{(a)}+n\to C^* \to Y^{(b)}+\gamma$, where  $C^*$ is a compound
nucleus with many closely spaced energy levels (high level
density). For the neutron star crust conditions the incident
neutron has rather low energy and the primary reaction mechanism
is dominated by compound formation. The partial cross section
$\sigma_{ab}(E)$ in the HF model is written as the sum over levels
$c$ (specified by energy $E_c$, spin $J$, and parity $\pi$) of the
compound nucleus
\begin{equation}\label{eq:sigma_HF}
  \sigma_{ab}(E)=\frac{\pi}{k^2} \sum_{c}
  \frac{g_c}{g_ng_a}
  \frac{{\cal T}_{n, a}^c {\cal T}_{\gamma, b}^c}{{\cal
  T}_{\text{tot}}^c}.
\end{equation}
In this case, $k$ is the wave number of an incident neutron,
 ${\cal T}_{n,a}^c$ and
${\cal T}_{\gamma,b}^c$
are partial transmission coefficients, and ${\cal T}_{\text{tot}}^c$
is the total transmission coefficient of the compound nucleus in a
level $c$.
The latter quantity, ${\cal T}_{\text{tot}}^c\equiv\sum_{o,b} {\cal
T}_{o,b}^c$, gives the total width of the $c$ level as the sum over
all available outgoing reaction channels $o=n,\, \gamma,\,
\text{etc.}$ and over levels $b$ of the final nucleus. Note, that the
sum includes compound elastic scattering (when final states are the
same as the initial ones).

The individual neutron transmission coefficient for each allowed
channel is obtained by solving the Schr\"odinger equation with an
optical potential for the neutron-nucleus interaction. The
$\gamma$-transmission coefficient  is calculated for a giant
dipole resonance (E1+M1) approximated by a single Lorentzian or by
a combination of Lorentizains \cite{ba1957,ku1990,junghans2008}.
The sum of these  contributions  determines the $\gamma$-ray
strength function.  Both the neutron and  $\gamma$-ray
transmission coefficients must be calculated for all accessible
states. In practice there is a huge number of levels, the vast
majority of which are experimentally unexplored. For very
neutron-rich nuclei near the drip line the level density may be
much smaller and the applicability of the HF model may be
questionable.

Other necessary ingredients for a HF calculation include the
choice of level density, optical model, $\gamma$-ray strength
function, and mass model to predict the reaction $Q$ value. The
reaction cross sections presented here do not include
pre-equilibrium effects. The cross sections are calculated on the
basis of $Q$ values derived from the Hartree-Fock-Bogoliubov mass
model HFB-17~\cite{goriely2009}. The level densities are obtained
from the microscopic model of Ref.~\cite{goriely2001}, and the E1
$\gamma$-ray strength function is based on quasiparticle
random-phase-approximation calculations, folded with a simple
Lorentzian~\cite{goriely2002}. The neutron optical potential is
supplied by the global parametrizations of Ref.~\cite{koning2003}.

\begin{widetext}
  \begin{figure}[th]
    \includegraphics[width=0.95\textwidth]{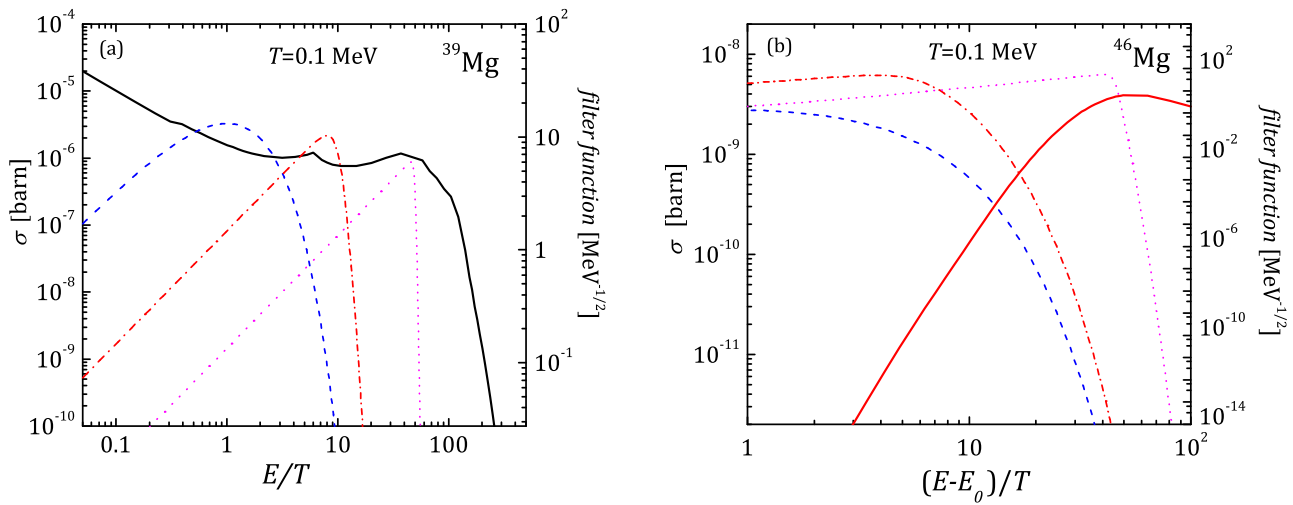}
        \caption{ Cross sections $\sigma$ of neutron capture
        on $^{39}$Mg (panel (a)) and $^{46}$Mg (panel (b))
        plotted (left vertical axis) in double logarithmic scale
        as a function of $(E-E_0)/T$
        at $T=0.1$~MeV. In both panels we show also (right vertical axis)
        the filter functions $E f(E)/{\cal N}$
        for the three cases: Maxwell-Boltzmann distribution (dashed
        lines) and Fermi-Dirac distribution with $\mu_{n}=1$~MeV (dash-dotted
        lines) and $\mu_{n}=5$~MeV (dotted lines). See text for details.}
    \label{fig:CrosSects}
  \end{figure}
\end{widetext}

Figure~\ref{fig:CrosSects} shows the reaction cross sections (left
vertical scales) for two neutron capture reactions, on $^{39}$Mg
(panel (a)) and $^{46}$Mg (panel (b)). The
$^{39}$Mg$(n,\gamma)^{40}$Mg reaction is exothermic ($E_0=0$),
while the $^{46}$Mg$(n,\gamma)^{47}$Mg is endothermic
($E_0=4.06$~MeV). For a better visualization of the approximation
(\ref{eq:sigma_pl}), $\sigma(E)$ is shown as a function of
$(E-E_0)/T$ on a double logarithmic scale; the temperature is
taken to be $T=0.1$~MeV. The linear segments of the curves clearly
indicate the power-law behavior of $\sigma(E)$ at low $E$.
Power-law indices and maximum energies are $\nu=-0.6$ and
$E_\mathrm{max}\approx 0.1$ MeV for neutron capture on $^{39}$Mg;
$\nu=3.5$ and $E_\mathrm{max}\approx 6$ MeV for neutron capture on
$^{46}$Mg. This power-law behavior at low energies is typical for
$(n, \gamma)$ reactions. Figure~\ref{fig:CrosSects} also shows
(right vertical scales) the so-called filter functions
$E\,f(E)/{\cal N}$, which enter the integrand of (\ref{eq:rate})
along with $\sigma^*(E)$. The dashed lines in both panels
correspond to the Maxwell-Boltzmann distributions of neutrons. We
see that the power-law approximation is sufficient for calculating
$\langle\sigma v\rangle_{\text{MB}}$ in both cases (at $T=0.1$
MeV). The dash-dotted and dotted lines in Fig.~\ref{fig:CrosSects}
represent the Fermi-Dirac distribution with $\mu_n=1$~MeV and
5~MeV, respectively. We see that for $\mu_n=1$~MeV and the
$^{46}$Mg target the power-law approximation is definitely valid,
while for $\mu_n=5$~MeV it is less accurate. For neutron capture
on the $^{39}$Mg nucleus, the power-law approximation is
inaccurate at both values of $\mu_n$.

It is easy to see that the power-law approximation
(\ref{eq:sigma_pl}) is valid as long as $\max (E_0,\mu_n)+T
\lesssim E_\mathrm{max}$. In this approximation, the factor $R_n$
in Eq.~(\ref{eq:R}) is calculated analytically. By introducing
dimensionless parameters $y=\mu_n/T$ and $x_0=E_0/T$, we obtain,
\begin{equation}
  \label{eq:R_powerlaw}
  R_n= \frac{\exp{x_0}}{x_0+\nu+1}\;
  \frac{(\nu+1){\cal F}_{\nu+1}(y-x_0)+x_0{\cal F}_\nu(y-x_0)}{{\cal
  F}_{1/2}(y)},
\end{equation}
where ${\cal F}_\nu(y)$ is a Fermi-Dirac integral
\begin{equation}
  \label{eq:IntFD}
  {\cal F}_\nu (y)= \frac{1}{\Gamma(\nu+1)}
  \int\limits_0^\infty \frac{x^\nu~\text{d}x}{1+\exp(x-y)},
\end{equation}
and $\Gamma(\nu+1)$ is the Euler gamma-function.

Equation (\ref{eq:R_powerlaw}) is the main result of our
consideration. We expect that this factor is sufficient to correct
the reaction rate for neutron degeneracy in many cases of
practical importance. Let us analyze the limiting cases. For this
purpose we use asymptotes of the Fermi-Dirac integrals given in
the Appendix.

First consider a threshold reaction with typical neutron energies
well below the threshold. In this case $x_0-y\gg 1$ ($E_0-\mu_n
\gg T$), and Eq.~(\ref{eq:R_powerlaw}) becomes
\begin{equation}
\label{eq:R_underthres} R_n=\frac{\exp y}{{\cal F}_{1/2}(y)}.
\end{equation}
If, in addition, neutrons are strongly degenerate ($y\gg 1$),
Eq.~(\ref{eq:R_underthres}) is further simplified,
\begin{equation}
\label{eq:R_deg_underthres}
  R_n=\frac{3\sqrt{\pi}}{4}y^{-3/2}\,\exp y.
\end{equation}
The factor $R_n$ in Eqs.~(\ref{eq:R_underthres}) and
(\ref{eq:R_deg_underthres}) becomes a function of only one
parameter, $y=\mu_n/T$; it is independent of $\nu$. This indicates
that Eqs.~(\ref{eq:R_underthres}) and (\ref{eq:R_deg_underthres})
are valid for any cross sections $\sigma(E)$ of threshold
reactions, not only for ones with power-law behavior. Indeed, in
the limit $x_0-y\gg 1$ the majority of neutrons have energies
$E<E_0$; these neutrons cannot overcome the reaction threshold and
cannot be captured by nuclei. The reaction proceeds owing to a
small amount of high-energy (suprathermal) neutrons with $E>E_0$.
Recall that for any neutron degeneracy, their distribution
function in Eq.~(\ref{eq:rate}) is actually Maxwellian,
$f_\text{FD}(E)\approx\exp\left(\mu_n/T\right) f_\text{MB}(E)$.
Then the cross sections in the nominator and denominator of
Eq.~(\ref{eq:R}) are integrated with the same function
$f_\text{MB}(E)$;  equal integrals for any $\sigma(E)$ cancel out
and do not affect the ratio $R_n$.

The ratio $R_n$ in Eq.~(\ref{eq:R_deg_underthres}) shows a sharp
exponential $y$-dependence at strong neutron degeneracy. This means
that neutron degeneracy exponentially enhances the rates of threshold
reactions (by increasing the amount of high-energy neutrons).

Another limiting case is the one of an (endo- or exothermic)
reaction with strongly degenerate neutrons whose typical energies
are above the reaction threshold. In this limit $y-x_0\gg 1$
($\mu_n-E_0 \gg T$), the reaction is driven by numerous energetic
Fermi-sea neutrons and becomes fast, with
\begin{equation}\label{eq:R_deg_overthres}
  R_n=\frac{3\sqrt{\pi}\exp x_0}{4\,\Gamma(\nu+3) }\;
  \frac{\left(y-x_0\right)^{\nu+1}}{y^{3/2}}\;\frac{x_0+(\nu+1)y}{x_0+\nu+1}.
\end{equation}

In particular, for a threshold reaction with strongly degenerate
neutrons at $E_0 \gg T$ and $\mu_n \gg T$, $R_n$ is given by
Eq.~(\ref{eq:R_deg_underthres}) for $E_0-\mu_n \gg T$ and by
Eq.~(\ref{eq:R_deg_overthres}) for $\mu_n-E_0 \gg T$. These two
asymptotes nearly match each other at $|\mu_n-E_0|\sim T$
providing an accurate description of $R_n$ in a wide range of
$\mu_n$ except for the narrow interval $|\mu_n-E_0|\lesssim T$
that should be described by Eq.~(\ref{eq:R_powerlaw}).

At $\mu_n>E_0$ the dependence of $R_n$ on the neutron degeneracy
parameter $y$ is much weaker than at $\mu_n<E_0$. In the range of
$T \ll \mu_n-E_0 \ll E_0$, Eq.~(\ref{eq:R_deg_overthres}) can be
simplified by setting $y=x_0$ everywhere but in $(y-x_0)^{\nu+1}$;
this gives $R_n \propto (\mu_n-E_0)^{\nu+1}$. For higher $\mu_n
\gg E_0$ we have $R_n \propto \mu_n^{\nu+1/2}$.

In order to calculate $R_n$ in intermediate cases from
Eq.~(\ref{eq:R_powerlaw}), accurate expressions for Fermi-Dirac
integrals are necessary. These integrals have been extensively
studied in the literature, especially in the field of
semiconductor physics and astrophysics (e.g., Ref.\
\cite{Blakemore82,cp98} and references therein). There are several
very accurate approximations for particular integer and
half-integer values of $\nu$. An analytic approximation that can
be used for any $\nu$ in the range $-1<\nu<4$ was constructed by
Aymerich-Humet et al.\ \cite{ahsmm82}. It accurately reproduces
the limits of $y\to\pm \infty$, and its relative error at
$-1/2<\nu<5/2$ does not exceed 1.2\%. For convenience, we present
this approximation in the Appendix.

\section{Plasma effects}
\label{S:plasma}

In addition to the effects of neutron degeneracy, the rates of
($n,\gamma$) reactions in dense matter are influenced by electron
plasma effects. Under typical conditions in the neutron star
crust, the electrons behave as weakly interacting, strongly
degenerate and ultra-relativistic particles \cite{hpyBOOK}. The
importance of plasma effects is characterized by the electron
plasma frequency $\omega_p=\sqrt{4\pi e^2 n_e/m^*_e}$, where $e$
is the electron charge, $n_e$ is the electron number density, and
$m^*_e=\mu_e/c^2$ is the effective electron mass at the Fermi
surface ($\mu_e$ being the electron chemical potential). The
plasma effects modify the radiative transition in the exit
channel. The plasma effects are strong when the frequency $\omega$
of a radiative transition becomes comparable to $\omega_p$. In
particular, no well-defined electromagnetic excitations (photons
or plasmons) can propagate at $\omega < \omega_p$ (e.g., Ref.\
\cite{abr84eng}). In this case the radiative transition cannot
occur through the emission of a real photon or plasmon. However,
it can occur through a direct transfer of the excess energy $\hbar
\omega$ to plasma electrons via collision-free collective
electromagnetic interactions. For a degenerate electron plasma in
the neutron star crust, this effect was considered in
Ref.~\cite{sy09}. It  does not suppress but rather enhances
radiative transitions at $\omega \lesssim \omega_p$. The
enhancement factor ${\cal R}_\gamma(\omega)$ depends on the
radiative transition type (electric or magnetic), on the
transition multiplicity $\ell=1,2,\dots$, and on the
$\omega/\omega_p$ ratio. In the low-frequency limit, $\omega\ll
\omega_p$, this factor behaves as ${\cal R}_{\gamma}(\omega)\sim
\left(\omega_p/\omega\right)^{2\ell}$, while at
$\omega\gg\omega_p$ one has ${\cal R}_\gamma(\omega)\to 1$,
meaning that the plasma effects become less important and the
standard regime of emission of real photons is restored.

The plasma effects modify the reaction cross section $\sigma(E)$ and
the reaction rate coefficient $\langle \sigma v \rangle$. In analogy
with Eq.~(\ref{eq:R}) we can formally introduce the total correction
factor
\begin{equation}
      R_{n,\rm pl}=\frac{\langle \sigma_\text{pl} v \rangle_\mathrm{FD}}
      {\langle \sigma v\rangle_\mathrm{MB}},
\label{eq:Rnpl}
\end{equation}
which takes into account both neutron degeneracy and plasma effects;
$\sigma(E)$ and $\sigma_\text{pl}(E)$ are the cross sections
excluding and including the plasma effects, respectively. It is
convenient to write
\begin{equation}
    R_{n,\rm pl}=R_n\,R_\mathrm{pl},
\label{eq:Rpl}
\end{equation}
where $R_n$ takes into account neutron degeneracy alone as discussed
in Sec.\ \ref{S:deg_n}, and $R_\mathrm{pl}$ accounts for plasma
effects (in the presence of neutron degeneracy).

The inclusion of plasma effects in  $(n,\gamma)$ reaction rates is
not straightforward. As discussed in Sec.~\ref{S:deg_n}, the rates of
interest are usually calculated \cite{rt00} in the framework of the
HF statistical model \cite{hf52}. The radiative transition
coefficients need to be modified in the numerator and denominator of
Eq.~(\ref{eq:sigma_HF}). The problem is further complicated by
summing over thermally excited nuclear levels $b$ in
Eq.~(\ref{eq:sigma_astro}). For different levels $b$ the energy and
type of radiative transition can be different (implying different
plasma modifications). A correct inclusion of the plasma effects is
therefore a complicated computational project, which is beyond the
scope of our paper. Here, we present a simplified approach which
demonstrates the importance of the plasma effects. It is based on the
assumption that the radiative transmission coefficients ${\cal
T}_\gamma$ are much smaller than other contributions to the total
transmission coefficient ${\cal T}_\text{tot}$ in
Eq.~(\ref{eq:sigma_HF}). In that case the total transmission
coefficient ${\cal T}_\text{tot}$ is independent of the plasma
effects. The second assumption is that the radiative exit channel is
represented by a single E1 radiative transition to the ground state,
so that no summation over excited states $b$ is required.

These requirements are realized for threshold ($n,\gamma$)
reactions involving degenerate neutrons where $\mu_n$ is below or
slightly above $E_0$. In this case the typical energy of
interacting neutrons is not much higher than $E_0$, which reduces
the radiative decay of the compound states to a single (not
multiple) low-energy radiative transition.

Under these assumptions we can describe the plasma effects by
multiplying the neutron capture cross section $\sigma(E)$ by the
factor ${\cal R}_{\gamma e1}(\omega)$, with $\hbar \omega=E-E_0$,
which describes the enhancement of the radiative transition by the
plasma effects. The latter factor was calculated in
Ref.~\cite{sy09} and fitted by a simple analytical expression
(Eq.~(33) in \cite{sy09}) with an error of about 1\%. The modified
cross section must be integrated over $E$ to obtain the reaction
rate. To simplify the integration we suggest using the
approximation
\begin{equation}\label{eq:Rgamma_simple}
  {\cal R}_{\gamma e1}(\omega)\approx 1+3.03 \left(\frac{\omega_p}{\omega}\right)^2.
\end{equation}
Although it is rather crude at $\omega\sim \omega_p$, with the
maximum error $17\%$ at $\omega=\omega_p$, it reproduces the
correct asymptotic behavior for $\omega\gg\omega_p$ and
$\omega\ll\omega_p$. The deviation does not exceed $5\%$ outside
the region of $0.6<\omega/\omega_p<1.03$. The advantage of using
Eq.~(\ref{eq:Rgamma_simple}) is that it allows an analytic
integration of the correction to the reaction rate in the same
power-law approximation for the cross section as used in
Sec.~\ref{S:deg_n}. Because ${\cal R}_{\gamma e1}(\omega)$ is
integrated, the approximation errors cancel out, leaving us with a
rather accurate result. The correction factor (\ref{eq:Rnpl}) to
the reaction rate including both neutron degeneracy and electron
plasma effects now becomes
\begin{equation}\label{eq:R_plasma}
  R_{n,\rm pl}=\frac{3.03}{\nu(\nu-1)} \left(\frac{\hbar \omega_p}{T}\right)^2
  \,\frac{x_0+\nu-1}{x_0+\nu+1}\; R_n^{(\nu-2)} +
  R_n^{(\nu)}.
\end{equation}
Here, $R^{(\nu)}_n$ is the factor given by Eq.~(\ref{eq:R_powerlaw})
for a power-law index $\nu$.

In the limit $E_0-\mu_n \gg T$  for a threshold reaction with $E_0
\gg T$, the expression (\ref{eq:R_plasma}) is further simplified.
Introducing the correction factor $R_\mathrm{pl}$  due to the
plasma effects in accordance with Eq.\ (\ref{eq:Rpl}), we have
\begin{equation}\label{eq:R_plasma_asymp}
  R_\mathrm{pl}=1+\frac{3.03}{\nu(\nu-1)} \left(\frac{\hbar
  \omega_p}{T}\right)^2 .
\end{equation}
In this limit, in addition to a strong exponential enhancement of the
reaction rate due to neutron degeneracy (Sec.\ \ref{S:deg_n}), there
is a smaller but significant plasma enhancement. In the opposite
limit of $\mu_n-E_0\gg T$ (but still for a single radiative
transition) we obtain
\begin{equation}\label{eq:R_plasma_overthres}
  R_\mathrm{pl}=1+3.03\, \frac{(\nu+2)(\nu+1)}{\nu(\nu-1)}
  \left(\frac{\omega_p}{\omega}\right)^2
  \frac{E_0+(\nu-1)\mu_n}{E_0+(\nu+1)\mu_n}.
\end{equation}
This factor is temperature independent because the typical transition
energy is now nearly fixed by $\mu_n$ and $E_0$, $\hbar \omega
\approx \mu_n-E_0$.

\section{Rates of inverse reactions}
\label{S:reverse}

If the rate of a forward $X(n,\gamma)Y$ reaction is known, then the
rate for an inverse reaction $Y(\gamma,n)X$ can be determined from
the detailed balance principle,
\begin{equation}\label{eq:reacbalance}
  n_{X}^\mathrm{(eq)} n_n^\mathrm{(eq)} \langle \sigma v\rangle = n_{Y}^\mathrm{(eq)}
  \lambda_\gamma,
\end{equation}
where $n_{X}^\mathrm{(eq)}$, $n_{Y}^\mathrm{(eq)}$, and
$n_n^\mathrm{(eq)}$ are number densities of nuclei $X$, $Y$, and
neutrons, respectively, in statistical equilibrium between forward
and inverse reactions. $\lambda_\gamma$ [s$^{-1}$] specifies the
rate of the inverse (photodisintegration) reaction (which is $n_Y
\lambda_\gamma$, cm$^{-3}$ s$^{-1}$). Usually, this reaction
involves only photons $\gamma$ but in our case it also can involve
more complicated excitations associated with the electromagnetic
field and plasma electrons (Sec.~\ref{S:plasma}), which are
assumed to be in thermal equilibrium; their effective number
density is included in $\lambda_\gamma$. The equilibrium number
densities of the nuclei should satisfy the condition of chemical
equilibrium
\begin{equation}\label{eq:chembalance}
  \mu_X+\mu_n=\mu_Y,
\end{equation}
where $\mu_X$ and $\mu_Y$ are the chemical potentials of the
nuclei $X$ and $Y$, respectively. Traditionally one assumes ideal
nondegenerate gas conditions for the nuclei and neutrons in order
to relate their equilibrium number densities and chemical
potentials. In this approximation Eq.~(\ref{eq:reacbalance})
yields the well-known relation
\begin{equation}\label{eq:detbalMw}
  \lambda_\gamma=\left(\frac{A_X
  m_n T}{2 \pi \hbar^2 A_Y}\right)^{3/2} \frac{2
  {\cal Z}_X}{{\cal Z}_Y}\; \exp\left(-{Q \over T} \right)
  \langle\sigma v\rangle_\mathrm{MB},
\end{equation}
where $A_X$ and $A_Y=A_X+1$ are  mass numbers of nuclei $X$ and $Y$,
respectively, while ${\cal Z}_X$ and ${\cal Z}_Y$ are their
individual internal partition functions
\begin{equation}
  {\cal Z}_X=\sum_a g_a \exp\left(-{E^{(a)}_X \over T}\right).
\label{eq:partition-fun}
\end{equation}
The corresponding function for the neutrons is  ${\cal Z}_n=2$.

Now we should modify Eq.~(\ref{eq:detbalMw}) to account for neutron
degeneracy and strong Coulomb coupling of the nuclei.

Strong Coulomb coupling prevents treating the plasma of atomic
nuclei as an ideal gas (it becomes Coulomb liquid or crystal;
e.g., Ref.\ \cite{hpyBOOK}). A strongly coupled multi-component
plasma of charged particles satisfies (to a very high accuracy)
the linear mixing rule according to which the main Coulomb
thermodynamic quantities (like mean Coulomb energy, etc.) can be
presented as sums of quantities for individual ions. Coulomb
coupling of an individual atomic nucleus $X=(A_X,Z_X)$ in this
plasma is described by the parameter $\Gamma_X=Z_X^2e^2/(a_XT)$,
where $a_X$ is the ion sphere radius defined as $a_X=\left[3
Z_X/(4\pi n_e)\right]^{1/3}$. This allows one to treat a strongly
coupled system of atomic nuclei as an ensemble of weakly
interacting ion spheres. This approximation is well known in the
physics of strongly coupled Coulomb plasmas \cite{hpyBOOK}. In
this case Eq.~(\ref{eq:detbalMw}) remains the same but the
internal partition function for each nucleus has to be multiplied
by its individual Coulomb partition function ${\cal Z}^{(C)}_X$. A
function ${\cal Z}^{(C)}_X$ depends only on one parameter,
$\Gamma_X$, which, in turn, is determined by the nuclear charge
number $Z_X$. Because in our case $Z_X=Z_Y$, we have ${\cal
Z}^{(C)}_X={\cal Z}^{(C)}_Y$, and the Coulomb corrections for the
nuclei $X$ and $Y$ compensate for each other in
Eq.~(\ref{eq:detbalMw}).

Neutron degeneracy can be included in Eq.~(\ref{eq:detbalMw})  by
implying the correct relation between the neutron number density
and its chemical potential. It is easy to show that for this
purpose it is sufficient to multiply the right-hand side of
Eq.~(\ref{eq:detbalMw}) by
\begin{equation}\label{eq:RDB}
  R_{\text{rvs}}=\exp(-y) {\cal F}_{1/2}(y).
\end{equation}
The ratio of the photodisintegration rates for the Fermi-Dirac and
the Maxwell-Boltzmann distributions of neutrons then becomes
\begin{equation}\label{eq:Rlambda}
  \frac{\langle\lambda_\gamma\rangle_{\text{FD}}}{\langle\lambda_\gamma\rangle_\text{MB}}\equiv
  R_\lambda=R_\text{rvs} R_n.
\end{equation}
If plasma effects are included, then $R_n$ must be replaced by
$R_{n,\rm pl}$.

Various asymptotes for $R_\lambda$ are readily obtained from those
for $R_n$, Eqs.~(\ref{eq:R_underthres})--(\ref{eq:R_deg_overthres}).
In particular, for the case of a threshold neutron capture reaction
with the neutron chemical potential well under the threshold,
$E_0-\mu_n \gg T$, we obtain (neglecting plasma effects)
$R_\lambda=1$ for any dependence of $\sigma$ on $E$. The inverse
reaction is not affected by neutron degeneracy which is quite
natural. The effect of neutron degeneracy on an ($\gamma,n$) reaction
consists of Pauli blocking of the emitted neutrons. However, in our
case these neutrons have energies $E=Q+\hbar \omega \gg\mu_n$, above
the Fermi level, where the blocking does not occur.

In the opposite limit, $\mu_n-E_0\gg T$, the neutrons, emitted in the
reverse reaction, have low energies and are strongly blocked by the
Fermi sea neutrons; this exponentially suppresses the inverse
reaction rate:
\begin{equation}
  \label{eq:Rlambda_overthres}
  R_\lambda= \frac{(y-x_0)^{\nu+1} \left[x_0+(\nu+1)y\right]}{(x_0+\nu+1)\,\Gamma(\nu+3)}\;
   \exp(x_0-y).
\end{equation}

Note that recently Mathews et al.~\cite{mathewsetal11} suggested
modifying the detailed balance equation (\ref{eq:detbalMw}) by
taking into account the quantum corrections due to induced photon
effects in the photodisintegration rate coefficient
$\lambda_\gamma$. They point out that while calculating
$\lambda_\gamma$ one usually employs the Maxwellian distribution
of photons instead of the Planck distribution $f_{\rm
Pl}=(\exp(E_\gamma/T)-1)^{-1}$. Using the Planck distribution,
they corrected $\lambda_\gamma$ and concluded that one should also
correct the detailed-balance ratio $\lambda_\gamma/\langle \sigma
v\rangle_{\rm MB}$. However, in this latter conclusion the authors
erroneously neglected the same corrections in the rate coefficient
$\langle \sigma v \rangle_{\rm MB}$ of the forward reaction.
Specifically, they did not include an extra factor $(1+f_{\rm
Pl}(E_\gamma))$ (with $E_\gamma=E+Q$) under the integral in their
Eq.~(6) (similar to Eq.~(\ref{eq:rate}) in the present paper) to
account for the induced emission. If that factor would have been
introduced, the detailed balance ratio would be naturally
unaffected by quantum corrections. In our analysis we neglect such
corrections in both forward and reverse reaction rates, because
they are generally small \cite{mathewsetal11}.

\section{Discussion}
\label{S:discuss}

\begin{figure}[ht]
  \includegraphics[width=0.8\columnwidth]{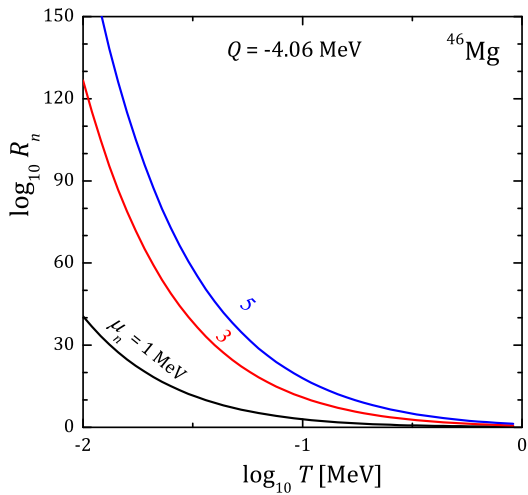}
  \caption{ Factor $R_n$ for the
  $^{46}\text{Mg}(n,\gamma)^{47}\text{Mg}$ reaction as a
  function of temperature for $\mu_n=1$, $3$, and $5$~MeV.
  }
  \label{fig:mg46}
\end{figure}

Let us illustrate the obtained results. First consider the effects of
neutron degeneracy on threshold reactions neglecting plasma physics
effects. Figure~\ref{fig:mg46} shows $R_n$ as a function of $T$ for
the $^{46}\text{Mg}(n,\gamma)^{47}\text{Mg}$ reaction at
$\mu_n=1,\,3$, and $5$~MeV. This reaction has a rather high
threshold, $E_0=4.06$ MeV. We see a strong increase of the reaction
rate with growing $\mu_n$. For $\mu_n=1$ and $3$~MeV the factor $R_n$
is well described with the power-law approximation by
Eq.~(\ref{eq:R_underthres}). For $\mu_n=5$~MeV the difference to the
power-law approximation becomes noticeable, but not on the
logarithmic scale of Fig.~\ref{fig:mg46}.

\begin{figure}[ht]
  \includegraphics[width=0.8\columnwidth]{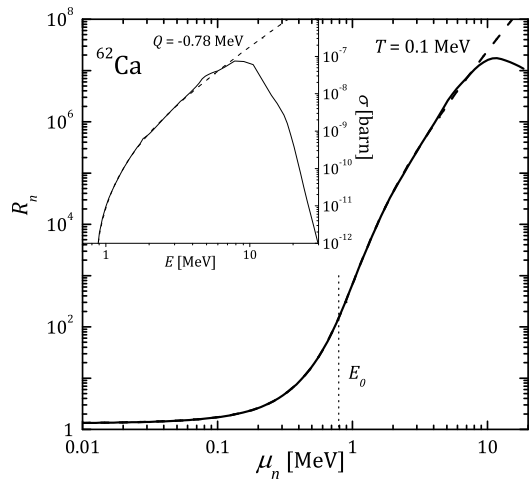}
  \caption{Factor $R_n$ for the $^{62}\text{Ca}(n,\gamma)^{63}\text{Ca}$
  reaction as a function of $\mu_n$ at $T=0.1$~MeV.
  Vertical dotted line indicates the reaction threshold.
  The inset shows the reaction cross section. Dashed lines in the
  figure and inset refer to the power-law approximation.
  }
  \label{fig:ca62}
\end{figure}

To illustrate the possible deviations from the power-law
approximation, in Fig.~\ref{fig:ca62} we plot $R_n$ versus $\mu_n$ at
$T=0.1$~MeV (the main figure) and $\sigma(E)$ (the inset) for the
$^{62}\text{Ca}(n,\gamma)^{63}\text{Ca}$ reaction
($E_0=-Q=0.78$~MeV). The solid lines are obtained by numerical
calculations, while the dashed lines are the results of the power-law
approximation. It can be seen that the latter approximation
accurately describes $\sigma(E)$ up to $E_\mathrm{max}\approx 10$
MeV. Accordingly, Eq.~(\ref{eq:R_powerlaw}) closely reproduces the
dependence of $R_n$ on $\mu_n$ at $\mu_n\lesssim 10$ MeV; the
exponential asymptote (\ref{eq:R_deg_underthres}) is valid at
$\mu_n\lesssim 1.5$ MeV; the power-law asymptote
(\ref{eq:R_deg_overthres}) works well at $1.5 \lesssim \mu_n\lesssim
$10~MeV. At $\mu_n\gtrsim 10$ MeV the power-law approximation becomes
inaccurate because at such $\mu_n$ the reaction rate is affected by
the high-energy segment of $\sigma(E)$ where the power-law is
invalid.

\begin{figure}[ht]
\includegraphics[width=0.8\columnwidth]{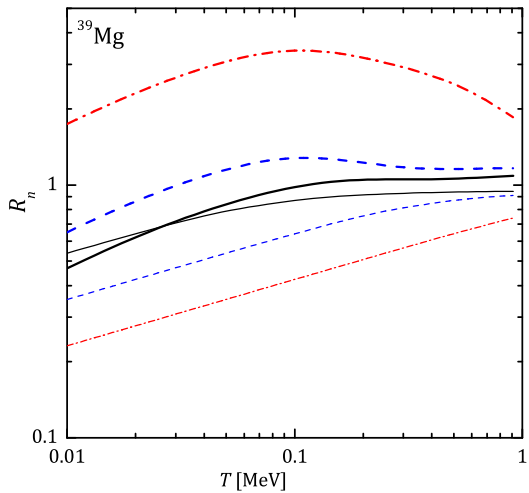}
\caption{ Factor $R_n$ for the
$^{39}\text{Mg}(n,\gamma)^{40}\text{Mg}$ reaction as a function of
temperature at $\mu_n=0.2$, $1$, and $5$~MeV (solid, dashed, and
dot-dashed lines, respectively). Shown are results of numerical
integration of calculated cross section $\sigma(E)$ (thick lines)
and results of the power-law approximation (thin
lines).}\label{fig:mg39}
\end{figure}

The situation is different with exothermic reactions. Figure
\ref{fig:mg39} shows the calculated factor $R_n$ for the
$^{39}\text{Mg}(n,\gamma)^{40}\text{Mg}$ reaction. The reaction is
exothermic ($E_0=$0, $Q=1.4$~MeV); its cross section is plotted in
the left panel of Fig.~\ref{fig:CrosSects}. The solid, dashed, and
dot-dashed lines in Fig.~\ref{fig:mg39} are calculated for
$\mu_n=0.2,\, 1$, and $5$~MeV, respectively. Thick lines are obtained
by integration of numerically calculated cross sections; thin lines
are obtained using the power-law approximation. We see that the
effect of neutron degeneracy is much weaker than for threshold
reactions; the factor $R_n$ stays $\sim 1$. Notice that for
$\mu_n=0.2$ and 1 MeV the Fermi-Dirac averaged rate is smaller than
the Maxwell-Boltzmann rate ($R_n<1$). The power-law approximation
breaks down even for small values of $\mu_n$ ($\sigma(E)$ starts to
deviate from power-law at $E\lesssim 0.2$~MeV, see
Fig.~\ref{fig:CrosSects}).

\begin{figure}[ht]
\includegraphics[width=0.8\columnwidth]{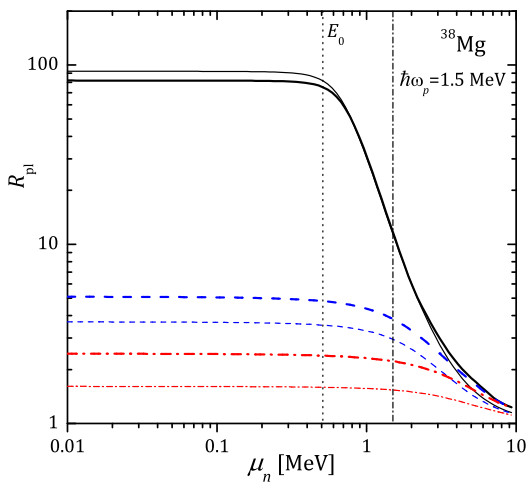}
\caption{ Factor $R_\mathrm{pl}$ vs $\mu_n$ for the
$^{38}\text{Mg}(n,\gamma)^{39}\text{Mg}$ reaction in matter with
the electron plasma frequency $\hbar\omega_p=1.5$~MeV. Solid,
dashed, and dash-dotted curves are plotted for $T=0.1$, $0.5$, and
$1$~MeV, respectively. Thick curves are obtained using computed
$\sigma(E)$ and accurate approximation of ${\cal R}_{\gamma
e1}(\omega)$; thin curves represent the approximation
(\ref{eq:R_plasma}). Vertical dotted and dash-dotted lines
indicate the reaction threshold and plasma frequency,
respectively.}\label{fig:mg38pl}
\end{figure}

Now let us discuss the impact of plasma effects on neutron capture
reactions starting with the threshold reactions. In all cases the
plasma effects cause an additional enhancement of the reaction rates.
Figure \ref{fig:mg38pl} gives $R_\mathrm{pl}$ for the
$^{38}\text{Mg}(n,\gamma)^{39}\text{Mg}$ reaction as a function of
$\mu_n$ at the electron plasma frequency $\hbar\omega_p=1.5$~MeV. The
curves are plotted for $T=0.1$ (solid lines), $0.5$ (dashed lines),
and $1$~MeV (dash-dotted lines). Thick curves are calculated with
numerically determined cross section and an accurate plasma
enhancement factor ${\cal R}_{\gamma e1}(\omega)$ from
Ref.~\cite{sy09}. Thin curves are given by simplified
Eq.~(\ref{eq:R_plasma}). All curves change their behavior when
$\mu_n$ reaches the reaction threshold $E_0=-Q=0.5$~MeV. For
$\mu_n<E_0$, the typical energy released in the radiative transition
is $\hbar \omega \sim T$, while at $\mu_n>E_0$ it is $\hbar \omega
\approx \mu_n-E_0$. Therefore, at $\mu_n<E_0$ the factor
$R_\mathrm{pl}$ is mostly independent of $\mu_n$, while at
$\mu_n>E_0$ it is independent of $T$ (cf.\
Eqs.~(\ref{eq:R_plasma_asymp}) and (\ref{eq:R_plasma_overthres})). A
significant plasma enhancement can be observed at $\mu_n<E_0$. It is
most visible for the solid curves (in those segments where $T$ and
$\mu_n$ are smaller than $\omega_p$). The plasma enhancement is
expected to be especially pronounced in those reactions which occur
in dense plasma and are accompanied by small radiative energy
release.

The difference between the thick (more accurate) and thin (less
accurate) lines in Fig.~\ref{fig:mg38pl} is small for $T=0.1$~MeV but
is higher for $T=0.5$ and $1$~ MeV. This is solely due to the
breakdown of the power-law approximation of $\sigma(E)$ for the
$^{38}\text{Mg}(n,\gamma)^{39}\text{Mg}$ reaction at $E\gtrsim
1.5$~MeV. We have checked that using Eq.~(\ref{eq:Rgamma_simple})
instead of the more accurate approximation to ${\cal R}_{\gamma
e1}(\omega)$ from Ref.\ \cite{{sy09}} does not significantly change
the results.

For exothermic reactions the simple model used in
Sec.~\ref{S:plasma} is generally inapplicable. We expect that
plasma effects on these reactions give $R_\mathrm{pl} \sim 1$.

\begin{figure}[ht]
\includegraphics[width=0.8\columnwidth]{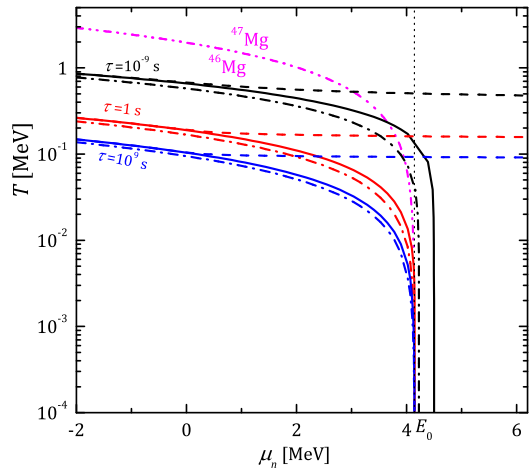}
\caption{ Lines of constant effective times $\tau=10^{-9},\,
1,\,10^9$~s of $^{46}$Mg burning in the
$^{46}$Mg$(n,\gamma)^{47}$Mg reaction on the $T$--$\mu_n$ plane.
We present calculations using the Maxwell-Boltzmann neutron
capture rate (dashed lines), the rate corrected for neutron
degeneracy (solid lines), and the rate corrected for degeneracy
and plasma physics effects  (dash-dotted lines). Vertical dotted
line positions the reaction threshold $E_0=4.06$~MeV. The
double-dot-dashed line corresponds to equal amounts of $^{46}$Mg
and $^{47}$Mg nuclei, assuming statistical equilibrium with
respect to neutron capture and emission reactions.}\label{fig:tau}
\end{figure}

Finally, in order to illustrate the efficiency of neutron captures,
in Fig.~\ref{fig:tau} we plot the lines of constant characteristic
burning times $\tau$ of $^{46}$Mg nuclei in the
$^{46}$Mg$(n,\gamma)^{47}$Mg reaction. The plot is made in the
$T$--$\mu_{\rm n}$ plane for the three values of $\tau=10^{-9},\,
1,\,10^9$~s. The characteristic burning time is defined as
\begin{equation}\label{eq:tau}
  \tau^{-1}= n_n\langle \sigma v\rangle.
\end{equation}
In the region above and to the right of each line burning is faster
than on the line ($\tau$ is smaller); in the region below and to the
left burning is slower. Lines of different types are calculated using
different reaction rates. The dashed lines are for non-degenerate
neutrons, the solid lines take into account neutron degeneracy, and
the dot-dashed lines take into account neutron degeneracy and plasma
effects (assuming $\hbar\omega_p=1.5$~MeV). Neutrons are essentially
non-degenerate in the region of $\mu_n \lesssim 0$ (in the left part
of the $\mu_n-T$ plane). They are strongly degenerate in the region
of $\mu_n \gtrsim T$.

As long as neutrons are non-degenerate, the solid and dashed lines
naturally coincide. When neutron degeneracy sets in, it intensifies
neutron captures. Then the solid lines (in sharp contrast with the
dashed ones) bend and become nearly vertical. In this regime neutron
degeneracy is vitally important. The most remarkable effect occurs in
the vicinity of the threshold ($\mu_n=E_0=4.06$~MeV, shown by the
dotted vertical line). We see that the solid lines drop off to zero
temperature almost immediately after $\mu_n$ exceeds the reaction
threshold. In this case the reaction is driven by the Fermi sea of
degenerate neutrons and becomes extremely fast.

The purpose of Fig.\ \ref{fig:tau} is primarily to illustrate the
efficiency of the $^{46}$Mg$\to^{47}$Mg transformation with
respect to neutron capture, neglecting other reactions (reverse
reaction, beta captures, and fusion reactions). Its main and
natural result is that $^{46}$Mg cannot survive for a long time
against neutron capture when $\mu_n$ exceeds $E_0$.

According to Fig.\ \ref{fig:tau}, the $^{46}$Mg$\to^{47}$Mg
reaction actually occurs in a narrow strip on the $T-\mu_n$ plane
that is confined between the lines $\tau=10^{-9}$ s and
$\tau=10^9$ s. Below and to the left of the $\tau=10^9$ line,
$\tau$ is very large; there will be plenty of $^{46}$Mg nuclei
which are very inefficient neutron absorbers. Above and to the
right of the $\tau=10^{-9}$ s line, $\tau$ is extremely short; all
$^{46}$Mg nuclei are transformed into $^{47}$Mg. The shape of this
``burning'' strip is similar to that for fusion reactions; see,
for instance, Fig.\ 4 of Ref.\ \cite{gasquesetal05}, which gives
the strip for the $^{12}$C+$^{12}$C fusion reaction in the
$T-\rho$ plane. The bend of the carbon burning $\tau$=const lines
at high densities $\rho$ is produced due to the transition from
thermonuclear carbon fusion to pycnonuclear one. It greatly
resembles the bend of the $\tau$=const curves with the growth of
$\mu_n$ due to the effects of neutron degeneracy in our Fig.\
\ref{fig:tau}.

In addition, the double-dot-dashed curve in Fig.\ \ref{fig:tau} is
the line representing equal amounts of $^{46}$Mg and $^{47}$Mg
nuclei ($n_X^{\rm (eq)}=n_Y^{\rm (eq)}$), assuming statistical
equilibrium with respect to the neutron capture and reverse
reactions. In statistical equilibrium, the matter (in our
simplified model) would mainly contain $^{46}$Mg nuclei below this
line and $^{47}$Mg nuclei above this line.

\section{Conclusions}
\label{S:concl}

We have considered  neutron captures ($n,\gamma$) in dense stellar
matter, taking into account the effects of neutron degeneracy and
plasma physics. The effects of neutron degeneracy increase the
amount of high-energy neutrons and mainly enhance the reaction
rates; plasma physics effect enhance the radiative transition in
the outgoing channel and enhance the reaction rates as well.

The effects of neutron degeneracy on neutron capture reaction
rates can be quantified by introducing the ratio $R_n$,
Eq.~(\ref{eq:R}), of rates calculated for given conditions to
those for non-degenerate neutrons. We have described this ratio by
a simple analytic expression (\ref{eq:R_powerlaw}) assuming the
power-law energy dependence of the reaction cross section
(\ref{eq:sigma_pl}) at energies that are not too high. The derived
expression (\ref{eq:R_powerlaw}) seems sufficient for many
applications. Furthermore, approximating $\sigma(E)$ by a
power-law function (\ref{eq:sigma_pl}), one also obtains the
power-law index $\nu$ and the maximum energy $E_\mathrm{max}$ to
which the power-law approximation of $\sigma(E)$ is valid. $E_0$
and $\nu$ are needed in Eq.~(\ref{eq:R_powerlaw}), while
$E_\mathrm{max}$ controls the validity of
Eq.~(\ref{eq:R_powerlaw}).

Our conclusions are as follows:

\begin{enumerate}

\item Neutron degeneracy can significantly affect ($n,\gamma$)
reactions in deep neutron star crust (Sec.\ \ref{S:deg_n}). In many
cases the effects of neutron degeneracy are well described by the
factor $R_n$ given by Eq.~(\ref{eq:R_powerlaw}). For threshold
reactions, strong neutron degeneracy enhances the reaction rate by
many orders of magnitude.

\item Plasma physics effects can additionally enhance ($n,\gamma$)
rates (Sec.\ \ref{S:plasma}), that is described by the factor
$R_\mathrm{pl}$. These effects are less dramatic but can reach a
few orders of magnitude.

\end{enumerate}

Furthermore, in Sec.\ \ref{S:reverse} we have used the detailed
balance principle and calculated the rates of inverse ($\gamma,n$)
reactions taking into account neutron degeneracy and plasma effects.
Finally, in Sec.\ \ref{S:discuss} we discussed the efficiency of
($n,\gamma$) reactions in a neutron star crust, with the conclusion
that neutron degeneracy can be most important.

Finally it should be noted that free degenerate neutrons in a
neutron star crust can be in a superfluid state. Critical
temperature $T_\mathrm{cn}$ for the appearance of neutron
superfluidity is very model dependent. Numerous calculations using
different techniques (e.g., Ref.~\cite{ls01}) give
density-dependent $T_\mathrm{cn}(\rho)$ with maximum values
ranging from $\sim 0.2-0.3$ MeV to $\sim 2$~MeV, indicating that
superfluidity is most likely. Superfluidity produces a gap in the
energy spectrum of neutrons near the Fermi level and modifies
matrix elements of neutron capture reactions. Both effects on
neutron captures are not explored but may strongly modify the
reaction rates.

Our consideration of neutron degeneracy and plasma effects on
neutron capture rates is simplified. A more rigorous (and
complicated) analysis of these effects (including also neutron
superfluidity) would be desirable. It would be instructive to
perform self-consistent calculations of the structure of atomic
nuclei immersed in a Fermi sea of free neutrons, taking into
account a compression of the nuclei by free neutrons (e.g., Refs.\
\cite{st83,hpyBOOK}). For simplicity, we have used a model of free
neutrons which occupy the space between atomic nuclei. At
densities $\rho$ not much higher than the neutron drip density, it
is sufficiently accurate (as follows, for instance, from results
of Ref.\ \cite{nv73}). In a self-consistent approach, this model
should be replaced by a more elaborated unified treatment of
neutrons bound in nuclei and free outside.

In any case one should bear in mind that neutron capture reactions
in a deep neutron star crust can be affected by neutron
degeneracy, plasma physics, and neutron superfluidity. These
effects may have important impacts on nuclear burning and
nucleosynthesis in the deep neutron star crust. The effects should
be taken into account to correctly simulate and interpret various
observational phenomena in accreting neutron stars such as X-ray
bursts and superbursts as well as quiescent thermal emission of
neutron stars in X-ray transients (e.g., Refs.\
\cite{sb06,schatz03,gu07} and references therein).

\begin{acknowledgments}
This work is partially supported by the Joint Institute of Nuclear
Astrophysics JINA (NSF-Phys-0822648). PSS and DGY acknowledge
support from RFBR (grant 11-02-00253-a), RF Presidential Program
NSh-4035.2012.2, and Ministry of Education and Science of Russian
Federation (contract 11.G34.31.0001). PSS acknowledges support of
the Dynasty Foundation. M.B. acknowledges the support from  the
ExtreMe Matter Institute EMMI in the framework of the Helmholtz
Alliance HA216/EMMI. DGY acknowledges support of RFBR (grant
11-02-12082-ofi-m-2011).
\end{acknowledgments}

\bibliography{NeutronStars}

\appendix*
\section{Approximation of Fermi-Dirac integrals by
Aymerich-Humet et al.}

The asymptotes of Fermi-Dirac integrals (\ref{eq:IntFD}) at $y\to
\pm \infty$ in nondegenerate and strongly degenerate limits are
easily obtained by retaining the first term in the series
expansion of the Fermi-Dirac distribution function over
$\exp(x-y)$:
\begin{subequations}
\label{eq:FD_asym}
\begin{eqnarray}
  {\cal F}_\nu(y) &=& e^{-y},\qquad y\to -\infty, \label{eq:FD_asymp_low}\\
  {\cal F}_\nu(y) &=& \frac{y^{\nu+1}}{\Gamma(\nu+2)},\qquad y\to
  +\infty.\label{eq:FD_asymp_high}
\end{eqnarray}
\end{subequations}

Aymerich-Humet et al.\ \cite{ahsmm82} derived a useful
approximation for Fermi-Dirac integrals, which, by construction,
reproduces the asymptotic limits (\ref{eq:FD_asym}). Their
approximation reads
\begin{equation}\label{eq:Fapprox}
 {\cal F}_\nu(y)=\left(\frac{\Gamma(\nu+2)2^{\nu+1}}
 {\left[b+y+(|y-b|^c+a^c)^{1/c}\right]^{\nu+1}}+e^{-y}\right)^{-1},
\end{equation}
where the fit parameters are
\begin{subequations}
\label{eq:Fapprox_params}
  \begin{eqnarray}
    a&=&\left[1+\frac{15}{4}(\nu+1)+\frac{1}{40}(\nu+1)^2\right]^{1/2},\label{eq:Fapprox_params_a}\\
    b&=&1.8+0.61\nu,\label{eq:Fapprox_params_b}\\
    c&=&2+(2-\sqrt{2})2^{-\nu}.\label{eq:Fapprox_params_c}
  \end{eqnarray}
\end{subequations}
In the range  $-0.9<\nu<4$ for any $y$ this approximation gives a
relative error of $\approx 1\%$. It is also valid for larger $\nu$
but become less accurate (to about $4\%$ at $\nu=12$).

\end{document}